\documentclass[twocolumn,showkeys,preprintnumbers,amsmath,amssymb]{revtex4}

\usepackage{graphicx}
\usepackage{dcolumn}
\usepackage{bm}

\begin{document}

\title{A chaotic dynamical reduction model for the quantum mechanical state vector}

\author{H. Brusheim-Johansson}
 \email{henbru-0@student.ltu.se}
\author{J.Hansson}
 \email{c.johan.hansson@ltu.se}
\affiliation{
Department of Physics\\{Lule\aa} University of Technology\\SE-971 87 {Lule\aa}, Sweden}

\date{\today}

\begin{abstract}
A new model is proposed for the purpose of modelling the ``wave function collapse'' of a two-state quantum system. The collapse to a classical state is driven by a nonlinear evolution equation with an extreme sensitivity to absolute phase. It is hypothesized that the phase, or part of it, is displaying chaotic behaviour. This chaotic behaviour can then be responsible for the apparent indeterminacy we are experiencing for a single quantum system. Through this randomness, the statistical ``ensemble'' behaviour, due to Born,  to describe a single quantum system, is no longer needed.
\end{abstract}


\maketitle

\section{Introduction}

The debate over the conceptual foundations of quantum mechanics has been going on since the theory emerged \cite{EPR1,EPR2,Jammer}. One aspect of the theory which so far remains to be solved is the measurement problem, i.e. how, when and why do the superposed quantum mechanical states turn into unique classical outcomes.\\
Although there certanly \emph{are} some viable non-collapse theories \cite{Everett1}, we will derive a novel non-local \cite{Bell1} model which explicitly reduces the statevector to an eigenstate. For an unobserved system, the state vector will evolve in a unitary (linear) way thus conserving superpositions, and we write the state vector as a (physical) expansion of a complete orthonormal eigenbase $\{\phi_n\}$. However as a consequence of observation, the statevector reduces, or projects onto one of the eigenstates in a \emph{nonlinear, nonunitary, irreversible} way;
\begin{equation}
|\Psi(t)\rangle=\sum_n c_n(t)|\phi_n(t)\rangle\longrightarrow|\phi_k(t)\rangle.
\label{eq:exp}
\end{equation}  
 In a dynamical theory like the present one, the projections in (\ref{eq:exp}) continue to evolve after a measurement in contrast to the ``standard'' interpretation. This will, in principle, provide us with the sought connection between the statistical ensemble-interpretation and the behaviour of an individual system. This nonlinearity associated with observation has to stem from some \emph{fundamentally} different type of quantum interaction than normally included in e.g. the Schr\"odinger equation.

 In general we never consider the direction of time in the laws of fundamental physics, instead, we just make some tacit assumption based on ``experience''. Herein lies the core of the reduction problem. In a reduction context, the direction of time is evident (irreversibility), but it is certanly \emph{not} retrievable from the standard (unitary) evolution, since it is by definition reversible. Irreversible processes are the rule in nature and reversibility is in most cases to be considered as an idealisation. We believe that to resolve the measurement problem, there is a need to incorporate instability, indeterminism and irreversibility into the governing equation since the Schr\"odinger equation is deterministic, timesymmetric and unitary in its present form, thus producing non-classical states as a consequence of its dynamics. So we realise that the Schr\"odinger equation describe an idealised stable world, certainly very accurate ``outside'' the inherently unstable reduction process, but not at all applicable at the moment of measurement.   
With this in mind, we see the need to modify the governing law of quantum mechanics in order to have an irreversible, unstable and time-assymetric description of nature on its fundamental level. We will pursue the idea that a chaotic behaviour is brought upon the system under measurement interaction, regardless of whether the reduction is induced by concious or inanimate physical systems. A chaotic evolution will lead to a dispersion of information over time. So we will, in practice, experience an irreversible evolution under measurement. This chaotic and unstable behaviour could then be the origin of the randomness we are experiencing on an individual level, for example the position of the hits on the detector in a Young double slit experiment. Our candidate for this chaotic behaviour is the quantum mechanical phase.\\
We consider the superposition as a highly unstable structure under measurement. So regardless of our initial condition, there are (usually) several states available to us. One of these states will get picked in a, for all practical purposes, indeterministic way due to the instability and chaotic evolution. \\
We will make the assumption that the Schr\"odinger equation in its most general form is governing the evolution at all times. However the exclusive measurement interaction is highly nonlinear and will destabilize the system (superposition) thus making it ``collapse'' on a characteristic time-scale, which must be small for macroscopic objects.\\

\section{Derivation of our model}
We have the following interaction situation for the Hamiltonians of the system before, under and after the interaction:
\begin{equation}
\mathcal{H}_0\Longrightarrow\mathcal{H_I+H}_0\Longrightarrow\mathcal{H}_0,
\end{equation}
 where $\mathcal{H}_0$ denotes the unobserved system and $\mathcal{H_I}$ is the interaction term.\\
Now, to incorporate an explicit phase dependance, we adopt the standard notation for projections in (\ref{eq:exp}) as 
\begin{equation}
c_n(t)=\sqrt{x_n(t)}e^{i\theta_n(t)},
\end{equation}
 where $\sqrt{x_n}\in[0,1]$. We acquire the time dependence of the probabilities ($x_n$) and phase ($\theta_n$) through the time dependence of the projections\begin{eqnarray}
\dot{\theta}_n&=&-\omega_n-\sum_m{\langle\phi_n|H_I^{NL}|\phi_m\rangle\sqrt{x_m x_n}\cos\left(\theta_m-\theta_n\right)}\nonumber\\
\dot{x}_n&=&\sum_m{\langle\phi_n|H_I^{NL}|\phi_m\rangle\sqrt{x_m x_n}\sin\left(\theta_m-\theta_n\right)}\label{xp}
.
\label{eq:pp}
\end{eqnarray} 
Now, it is evident that the matrix elements of the interaction Hamiltonian determines the collapse evolution completely. The only remnant of the ``unobserved'' Hamiltonian in the equations are the eigenenergies ($\hbar\omega$) \cite{Sakurai1, Pearle1}.\\
The nature of the interaction Hamiltonian has to be such that it drives all $x_n$ to zero, except one, say $x_k$, which is driven to unity. This gives us the connection between the superposition and the individual behaviour. Many previous approaches include the postulate
\begin{equation}
\dot{\theta}_n=0,
\end{equation}
 at least during the very brief time of interaction. This is obviously a very convenient approach. The key idea of this paper is, as mentioned above, the opposite. That is, the phase, or a part of it, is \emph{anything} but well behaved during the interaction, and fluctuates so violently that it is completely indeterminable (in practice), i.e., chaotic.\\
 It is not entirely clear just how this chaotic phase should act. We see two possibilities: either the phase acts \emph{commonly} for all states, or it will act individually on the different states in (\ref{eq:exp}). This should be determinable (in principle) in experiments where one looks for ``anomalies'' in high-resolution interference experiments.
In this interaction scheme, we do not see any possibility of a Herimitian interaction. Not least due to the consequence of an unitary evolution, which cannot describe a reduction process. The non-Hermiticy will however present new problems, since we have
\begin{equation}
\frac{d}{dt}\langle\Psi|\Psi\rangle=i\langle\Psi|\mathcal{H}^\dagger-\mathcal{H}|\Psi\rangle,
\end{equation}  
 i.e., norm conservation and Hermiticity are interconnected, at least in a \emph{linear} theory. With the hypothesis that the states correlate through the phase and there being \emph{no} correlation among the amplitudes, we propose the simplest possible equation for the probabilities
\begin{equation}
\dot{x}_n=f_n(\alpha_k)\alpha_nx_n(1-x^2_n).
\label{eq:prob}
\end{equation}
Where $\alpha_n=\frac{cos(\theta(0)_n)}{|cos(\theta(0)_n)|}$ is introducing an extreme instability with regards to phase into the system. In the present theory, the phase ``acts'' on the probability equation (\ref{eq:prob}) as long as the states are uncollapsed. However, once the collapse has been induced, the phase can no longer affect the evolution of the probabilities. Much like sneezing on a pen balancing on its tip will make it tip over, but sneezing on a falling pen cannot stop it from falling. So $t=0$ in the $\alpha$ function is to be interpreted as the moment when the collapse is (irreversibly) induced and the system (particle) is realised from the many potentialities in the wavefunction. The function $f_n(\alpha_k)$ is the coupling between states, or Einstein's infamous ``spooky action at a distance'' \cite{EPR1}, inherent to a non-local theory as the present one. It provides the necessary correlation through the phase.

\begin{widetext}
\begin{equation}
f_n(\alpha_k)=\left[1-2\sum_{k\neq n}\Theta_+\left(\alpha_k\right)\right]\cdot\alpha_n+\left[1-\sum_{k\neq n}\Theta_+\left(\alpha_k\right)\right]\cdot\Theta_+\left(\sum_{k\neq n}1-\Theta_+(-\alpha_k)\right)\cdot\left[1-\alpha_n\right].
 \label{eq:ad}
\end{equation}
\end{widetext}

Solving (\ref{eq:prob}), we get
\begin{equation}
x_n(t)=\frac{1}{\sqrt{1+\frac{1-x^2_n(0)}{x^2_n(0)}e^{-f(\alpha_k)\alpha_n t}}},
\label{mod}
\end{equation}
here, the $\Theta_+$'s are Heaviside functions, such that $\Theta_+(0)=0$.\\
To visualize the reduction process in a simple way, we turn to a quantum mechanical system with only two states (e.g. a spin 1/2 system) and form $q=x_1-x_2$, which will thus collapse to $\pm1$, depending on which state is selected.\\
\begin{figure}[hbt!]
\begin{center}
\includegraphics[scale=0.2,angle=270]{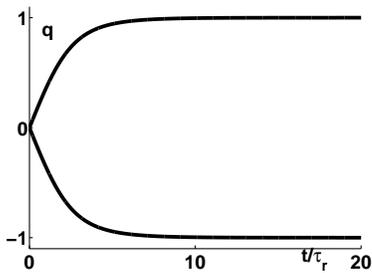} 	
\end{center}
\caption{\label{transient}Collapse of a two-state wave function, $\tau_r$=characteristic reduction time. A collapse to q=1 corresponds to a reduction of the state vector to $|\phi_1\rangle$, and a collapse to q=-1 corresponds to a reduction to $|\phi_2\rangle$. The strength of the interaction is inversely proportional to the reduction time. The transient part is open for potential experimental detection.  }
\end{figure}
Trying to get a rough estimate of the characteristic size of the reduction time as defined in the present theory, we need to look for some typical energy or interaction strength. Suggestions have been made \cite{Hansson} connecting the reduction time to non-linearities already present in the gauge-fields. This would give us an, at least qualitative, notion of the magnitude of the reduction time, since the interaction strength and thus the reduction time is, at least in principle, retrievable from the interaction energy stored in the gauge-field equations.\\
The chaotic phase part could possibly be detected in a high-resolution interference experiment. Anomalies should then be detected in the otherwise smooth interference pattern. However it would requrie a virtually noise-free experimental setting. As pointed out before, the dynamical reduction would have a distinct characteristic \cite{Costas1}, i.e. a smooth, continous behaviour rather than the instantaneous reduction prescribed by the orthodox quantum mechanics. Once again the resolution has to be extreme, making the technological difficulties great. 

\section{Conclusion and discussion}
We have pointed out the possibility for a quantum mechanical collapse process where the total quantum evolution is governed by only \emph{one} evolution equation, in contrast to two, Schr\"odinger's equation and Born's collapse postulate, in orthodox quantum mechanics. However, two fundamentally different interactions are needed. The dynamics proposed here contains nonlinear and chaotic terms introduced through a non-Hermitian Hamiltonian. Moreover the statistical behaviour is reproduced through the concept of indeterminable or chaotic absolute phase. Also, we propose that it is the phase that provide us with an ``EPR telephone'', i.e., the action at a distance, which is inherent in a non-local theory. This action is mediated through the phase, which thus in a sense has to work on a level ``above'' Minkowski space-time. Thus nonlinear and chaotic terms could be the remedy for the inability of ``orthodox'' quantum mechanics to describe the individual behaviour of a system. Using the phase as a specific kind of ``hidden variable'' has the advantage of not introducing any new and exotic variables into the theory. Nor have we postulated any extra evolution equations working in parallel with the usual equations $(x_n, \theta_n)$.\\
However, obtaining an appropriate $\tau_r$ is not so straightforward. Clearly, the interaction coupling determines the strength of the nonlinear term, or to put it another way, determines the speed of the collapse process. As such it obviously must be closely connected to the very physical process that induces the collapse. The process cannot be ``too slow'', because that would imply the ability to observe superpositions in the ``classical'' world (non collapsed states). The range of $\tau_r$ is such that it is small for large (or strongly interacting) systems and large for small (or weakly interacting) sytems. There has been suggestions of how to experimentally set some boundaries on $\tau_r$ \cite{Costas1}, and looking for the physical origin of the non-linearities, we could, at least in principle, make theoretical predictions of the magnitude of the reduction time (e.g. \cite{Hansson}).

\bibliography{apssamp}

\end{document}